\documentclass[titlepage]{amsart}
\usepackage{amsaddr}
 
\usepackage{booktabs} 
\usepackage{float}

\usepackage[pdftex]{graphicx}

\usepackage{url}
\usepackage{flushend}

\begin{document}

\title[SurfBraid]{SurfBraid: A concept tool for preparing and resource estimating quantum circuits protected by the surface code}

\author{Alexandru Paler}
\address{Linz Institute of Technology\\Johannes Kepler University, 4040 Linz, Austria}
\email{alexandrupaler@gmail.com}

\begin{abstract}
The first generations of quantum computers will execute fault-tolerant quantum circuits, and it is very likely that such circuits will use surface quantum error correcting codes. To the best of our knowledge, no complete design automation tool for such circuits is currently available. This is to a large extent because such circuits have three dimensional layouts (e.g. two dimensional hardware and time axis as a third dimension) and their optimisation is still ongoing research. This work introduces SurfBraid, a tool for the automatic design of surface code protected quantum circuits -- it includes a complete workflow that compiles an arbitrary quantum circuit into an intermediary Clifford+T equivalent representation which is further synthesised and optimised to surface code protected structures (for the moment, braided defects). SurfBraid is arguably the first flexible (modular structure, extensible through user provided scripts) and interactive (automatically updating the results based on user interaction, browser based) tool for such circuits. One of the prototype's methodological novelty is its capability to automatically estimate the resources necessary for executing large fault-tolerant circuits. A prototype implementation and the corresponding source code are available at \url{https://alexandrupaler.github.io/quantjs/}.
\end{abstract}

\maketitle

\section{Introduction}

Quantum circuit fault-tolerance is necessary due to the imperfect nature of quantum hardware. Fault-tolerance is mainly achieved by using a quantum error correcting code (QECC) \cite{NC00}, underlying the functionality of a quantum circuit synthesised from the corresponding quantum algorithm. Synthesised (compiled) circuits are applications of error corrected quantum gates on error corrected qubits. It is increasingly believed that for a QECC to be practical, it must have a error threshold of around 1\%, and under this constraint arguably the most practical choice is the surface QECC \cite{FMM13}, a topological code. Although quantum circuits based on topological codes require large numbers of physical resources (gates and qubits) for encoding and executing the computation, those numbers are realistically smaller than for any other quantum error correcting code possessing a similar threshold.

\begin{figure*}[ht!]
\centering
\includegraphics[width=0.8\textwidth]{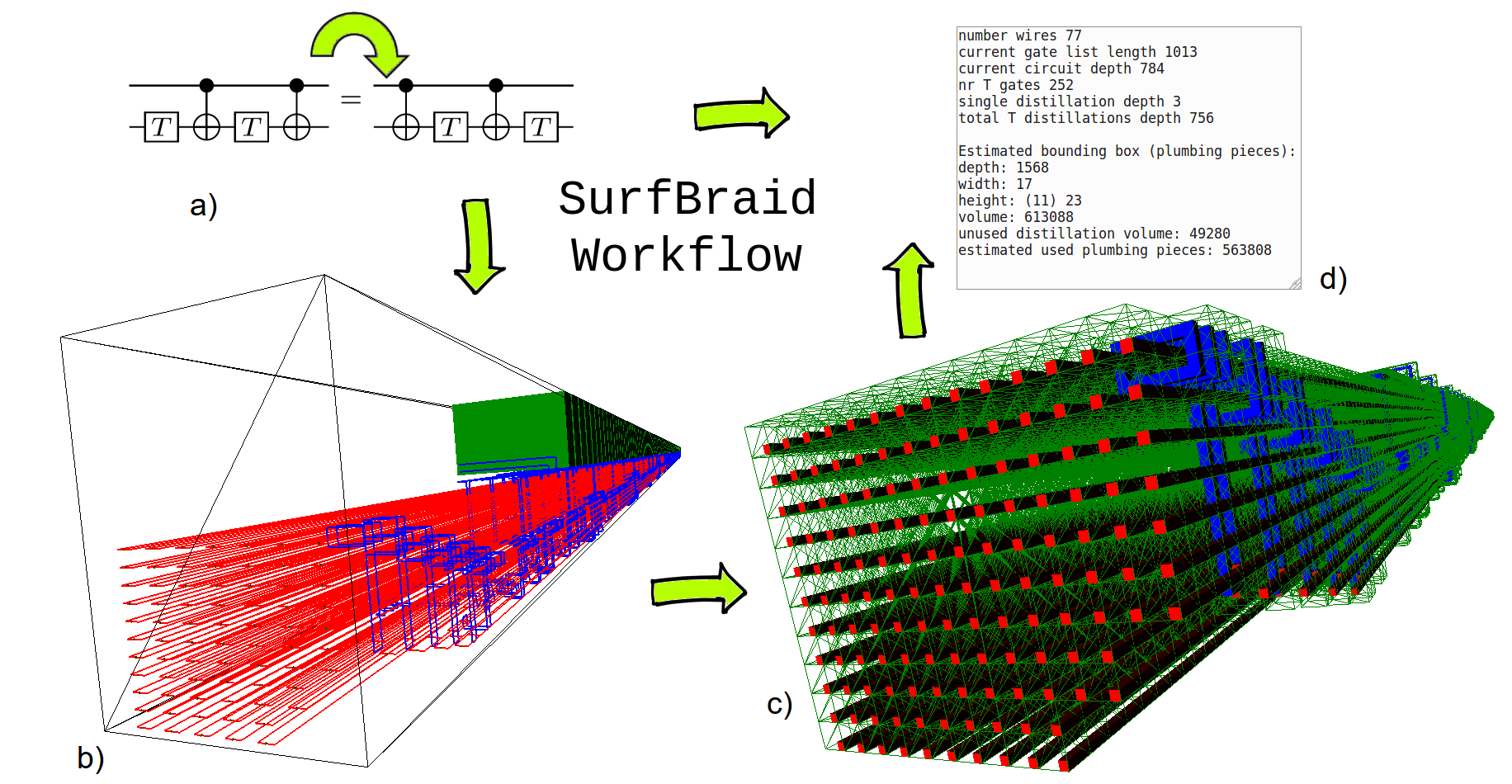}
\caption{SurfBraid workflow. Green arrows indicate existing functionality. The goal is to perform the a)-c) transformations and to generate analysis results similar to d) at each workflow step. a) Example of Clifford+T quantum circuit. The two circuits are computationally equivalent. A rewrite rule is the procedure that replaces one of the circuits with the other. b) Surface code layout including qubits (red lines), CNOT gates (blue lines), boxes (green) abstracting special quantum state preparation routines, and the bounding box (black lines) expressing the resources estimated to layout the circuit; c) A more detailed illustration of the resources required to layout the circuit from b) after leaving the green boxes out; d) analysis results example. (The views from c) and d) are in the \emph{Defects} and \emph{Plumbing} tabs of SurfBraid).}
\label{fig:qcirc}
\end{figure*}

Fault-tolerant quantum circuits using the surface code are traditionally restricted to Clifford+T gates \cite{FMM13}. The Clifford gate set is classically efficiently simulateable and is generated by the single qubit H, S gates and the two qubit CNOT gate \cite{NC00}. Quantum computational universality is typically achieved by augmenting the Clifford set with the T gate (e.g. Figure~\ref{fig:qcirc}a uses Clifford+T gates). It has been shown that Clifford+T gates can be decomposed into circuits consisting entirely of single qubit initialisations, CNOT gates, and single qubit measurements (performed in a well defined order) \cite{paler2017fault}.

\subsection{Surface Code Compilers and Resource Estimation}

Surface quantum error correcting codes have been intensively researched during the last decade (from \cite{raussendorf2007topological} to \cite{FMM13} and recently e.g. \cite{litinski2018game}), and form the basis of some of the most promising large scale quantum computing architectures (e.g. \cite{van2016path}). The focus on design automation frameworks for quantum circuits has increased, and this was motivated by the availability of NISQ hardware, such as the IBM chips, Rigetti computers and Google prototypes. Companies and research groups started developing various tool chains (e.g. \cite{steiger2018projectq}, \cite{cross2018ibm} or \cite{roetteler2017design}) for automating the preparation and execution of arbitrary quantum circuits to quantum machines (for the moment mostly noisy intermediate-scale quantum -- NISQ \cite{preskill2018quantum}). Design automation for large scale error corrected quantum computers has been neglected for many years, but is gradually being discovered by the research community. Nevertheless, to best of our knowledge, a tool able to automatically design surface code circuits has not been available until now. While a prototype was presented in \cite{paler2017synthesis}, its applicability was very limited, because it was difficult to feed it with arbitrary quantum circuits, control its execution, observe and optimise the generated surface code layouts.

Tools for the resource estimation of fault-tolerant quantum circuits have been proposed in the past, too. One of the first was \cite{suchara2013qure}, and other works such as \cite{lin2015paqcs} followed. Their designs and functionality were clear, but lacked flexibility. After Clifford+T optimisation established as a research topic per se, some works started discussing the cost of protecting various distillation procedures by the surface code (e.g. \cite{fowler2013surface}). More recently, the works of \cite{javadi2017optimized}, \cite{ding2018magic} treated the preparation and resource estimation of surface code protected quantum circuits. Although these tools are of very high quality, they do not consider the dynamic arrangement of the distillation procedures, are not interactive, and most of the times do not employ scalable heuristics or do not use high performance data structures for speeding up the preparation. Nevertheless, almost all of the previously mentioned tools are more than concepts, but they are difficult to extend. SurfBraid is built around a straightforward architecture such that it can be both flexible, scalable, interactive and offer as exact as possible resource estimations.

\subsection{Structure}

In the following, the implementation of the SurfBraid concept is presented. The tool is a working prototype whose functionality is being improved. The Results section briefly sketches some of the technical details underlying the concept. The practical performance of SurfBraid is shortly discussed in the same section. The Methods section enumerates the core features of the tool. The online (interactive) methodology is the sum of previous more theoretical works which focused on surface code preparation. The final section is an optimistic outlook on the future development of SurfBraid.

\section{Results}

A scalable (see Section~\ref{sec:perf}), online working alpha version of SurfBraid has been implemented at \url{https://alexandrupaler.github.io/quantjs/}. Source code is available at \url{https://github.com/alexandrupaler/quantjs/}.  The Surfbraid workflow is sketched in Figure~\ref{fig:qcirc} and includes the preparation (compilation and optimisation), the visualisation, and the resource estimations of surface code protected circuits. In the following these circuits will be called \texttt{scqc}.

\subsection{Technical Details}

SurfBraid is a \emph{(semi-)automatic} tool, in the sense that users can modify and control the currently implemented work flow. This allows to \emph{fine tune the parameters of compilation and optimisation} of \texttt{scqc}. Javascript was chosen for implementing SurfBraid, and this leads to an \emph{on-device computation} model, which has the advantage that it maintains data privacy and low latency \cite{smilkov2019tensorflow} (e.g. compared to cloud hosted applications such as the straightforward IBM circuit designer). Additional benefits of Javascript are the vast libraries ecosystem and standardisation of WebGL. These minimise the difficulty of portable source code maintenance, but are also very effective for implementing the interactive user experience.  The user experience is a pragmatic objective, although not of central methodological importance.

Interactive analysis and visualisation exists for both: 1) quantum circuits (by integrating Quirk\footnote{\url{http://algassert.com/quirk}} - which is also based on Javascript and WebGL), and for 2) the three-dimensional layouts of \texttt{scqc}. The intrinsic graphic nature of \texttt{scqc} is exploited for \emph{experimentation with surface code protected quantum circuits}. Quantum algorithm/circuit modifications are visible immediately (online).

\begin{figure}[t]
    \centering
    \includegraphics[width=0.4\textwidth]{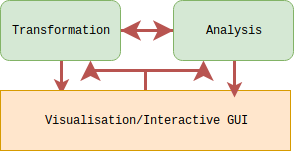}
    \caption{The three types of SurfBraid components and possible interactions between them. Each component can be used to receive inputs from another type of component. For example, a Clifford+T transformation component can send the output circuit to an analysis component that counts the number of resulting T gates. The interactivity of the tool implies that GUI components can steer the transformation and analysis components (e.g. a click transforms a circuit, whose analysis results are visualised dynamically.)}
    \label{fig:compinteraction}
\end{figure}

From a software architecture point of view, the tool includes loosely coupled components which interact through a minimal API. Component interactions are designed to behave similar to a Unix pipeline: each component takes as input a list of strings (e.g. list of quantum gates/instructions) and outputs another list of strings (e.g. two numbers indicating the number of T gates and the depth of the circuit). In general, the architectural design is similar to the one from ProjectQ\cite{steiger2018projectq}, but SurfBraid is less elegant, in order to be more maintainable -- until the entire tool architecture will stabilise and be more robust and resilient to software bugs. In particular, SurfBraid consists of three types of functionally different software components (Figure~\ref{fig:compinteraction}):
\begin{enumerate}
	\item \textit{transformation}
    \item \textit{analysis}
    \item \textit{visualisation}
\end{enumerate}

The simplistic API (list of strings are component inputs, and component outputs are strings, too), together with the three component types form the backbone of a \emph{plugin system}, which enables the tool's flexibility. The plugin system can interface with higher level languages (e.g. Quipper \cite{green2013quipper}) or lower level surface code decoders (e.g. for very small scale circuits Decodoku \cite{wootton2017decodoku}).

Each component type interprets some of the string inputs as parameters to its functionality. For example, a component that generates the gate list of an adder will take as parameter the number of qubits the adder operates on. The SurfBraid GUI adapts automatically to the supported parameters of the components. Thus, new components can be graphically operated when introduced into SurfBraid, increasing as a result the tool's interactivity.

\subsection{Workflow}

The tool's operations starts from an initial circuit which can be either generated or imported. These operations imply the usage of transformation components. More specifically, these components (called \emph{generators}) either: a) take string parameters as inputs (e.g. number of qubits, maximum depth etc.) and output a gate list (Clifford+T, or any other gate set specified by the user); or b) import strings representing files generated by other quantum circuit design tools. For the latter option, SurfBraid includes a generator that can interpret the file format from RevKit \cite{soeken2012revkit}. Due to this feature, a very large variety of reversible/quantum circuits can be imported directly.

In general, \emph{transformation} components are used for the translation of a quantum circuit format into another format. Consequently, SurfBraid takes the list of strings output by a generator and sends it to future transformations (e.g. Clifford+T to ICM \cite{paler2017fault}), or to analysis and visualisations components. \emph{Analysis} components extract quantitative information from a quantum circuit representation (e.g. T count).

\emph{Visualisation} components function as terminals (WebGL, PDF etc.) which receive commands/instructions from corresponding transformation components. For example, in the current version, a JSON file is generated by a transformation component and sent to the WebGL visualisation of \texttt{scqc}. A \texttt{scqc} three dimensional WebGL rendered visualisation example is Figure~\ref{fig:qcirc}b), where the green \emph{boxes} abstract the repeating surface code layout of a specific sub-circuit (called distillation circuit). The interactive GUI and the online workflow update visualisations on the fly: each modification is automatically visualised in an updated \texttt{scqc} layout. 

Figure~\ref{fig:qcirc} shows that after a circuit is generated, its gate list can be manually or automatically optimised and transformed into \texttt{scqc} layouts. This optimisation procedure is based on the iterative execution of a loop consisting of analysis, followed by transformation components. The loop includes the methods from Section~\ref{sec:distrib}, Section~\ref{sec:control}, and Section~\ref{sec:matching}.

Quantum circuits protected by the surface code (also referred to as \emph{topological assemblies}) are generated by a component compiled from the C++ code of \cite{paler2017synthesis} through Emscripten (an LLVM to Javascript compiler\footnote{https://github.com/emscripten-core/emscripten}). The entire functionality from the C++ version of \cite{paler2017online} is available in SurfBraid, but for the moment only some features have been unlocked in practice (in the GUI). The Emscripten compiled module is working as a transformation component, and its output is saved to in-memory text files, which are processed by transformation, analysis, and visualisation components.
 
It is desirable for SurfBraid to support a \emph{high degree of correctness for the generated results}. This can be achieved by running verification processes in background (e.g. \cite{zulehner2017advanced}), or by allowing the users to perform only correct operations. An example for the latter situation is the manual application of previously verified (known to be correct) rewrite rules \cite{maslov2005quantum} (e.g. Figure~\ref{fig:qcirc}a). The application of template rewrites is supported (see next Section~\ref{sec:matching}). From a Javascript perspective, future work will look at how verification could be sped up by using Tensorflow as presented in \cite{smilkov2019tensorflow}.

\subsection{Performance}
\label{sec:perf}

Detailed performance evaluations are not within the scope of this work because of two reasons: a) the tool is continuously improved and developed; b) existing empirical evidence shows that its performance is sufficient for state-of-the-art resource estimations -- for example, it was used to resource estimate the circuits from \cite{babbush2018encoding}.

Javascript has a performance penalty compared with lower-level programming languages, but compared with other interpreted languages the penalty is negligible or does not even exist. Contrary to common belief, Javascript is fast when executed in modern browsers or in Node.js\footnote{https://github.com/nodejs/node}. The majority of the tools for quantum circuit design is based on Python and their performance stems only from the use of \texttt{numpy}. SurfBraid can automatically design circuits of hundred of qubits, because it uses internally some non-trivial data structures \cite{paler2018faster} to significantly speed up common circuit level operations.

The empirical evidence is based on a resource estimation procedure (very similar to Figure~\ref{fig:qcirc}) which includes the steps: a) procedural generation of the QROM and Majorana circuits \cite{babbush2018encoding}, b) automatic template matching for the optimisation of the Clifford+T circuits, and c) the generation of the \texttt{scqc} layouts. Medium scale circuits (e.g. quantum adders of 64 qubits) fully error corrected by the surface code can be easily prepared and visualised on a common machine. The performance bottleneck is the \texttt{scqc} visualisation and not the Clifford+T decomposition, layout generation or computational resource estimation. For example, the resource estimation for the error corrected layouts of the circuits (up to 2000 logical qubits) from \cite{babbush2018encoding} can be performed within minutes on an i5 machine. These performance results were obtained by disabling visualisation, but still running SurfBraid in the browser.

SurfBraid is sufficiently fast to be of practical use to researchers. For very large quantum circuits, and circuits where visualisation of the layouts is not necessary, the source code could be executed in Node.js environments. Future work will include performance tweaks of the tool.

\section{Methods}

The central application of SurfBraid is to {\bf estimate the computational resources} necessary to execute an arbitrary \texttt{scqc} represented as a layout of braided defects (e.g. Figure~\ref{fig:qcirc}d)). The current version uses \emph{plumbing pieces} \cite{fowler2012bridge} as units for expressing the resource overheads.

Resource estimation are executed online: once the user changes a parameter in the GUI (e.g. number of qubits an adder is operating on), the resource estimations are automatically updated. Estimations are based on a single layout template similar to the one from Figure~\ref{fig:qcirc}c): the green boxes (distillation sub-circuits \cite{FMM13}) are placed in a row parallel to the qubits which are arranged to fit exactly below the boxes. To the best of our knowledge, this is the first automated tool able to perform such approximations.

SurfBraid is configurable with respect to the three dimensions of the green distillation boxes. This is necessary, because distillation circuits (and their corresponding green bounding boxes) are continuously reduced/improved (e.g. \cite{gidney2018efficient}), such that preparing and resource estimating \texttt{scqc} is a job that has to be repeated.
 
Additionally, SurfBraid supports saving the intermediate steps, such that a undo/redo mechanism is available. Almost all outputs of the components can be saved and downloaded (e.g. gate lists, three-dimensional structures).

\subsection{Two Gate Types}
\label{sec:gatetypes}

SurfBraid is agnostic of the gate types (e.g. H, S or Toffoli). This is because of the component model, where inputs and outputs are just strings. The component internal logic dictates how the strings are interpreted. Therefore, a quantum circuit gate list is just a sequence of strings, and a corresponding component treats all entries of the list as representations of two types of gates: 1) unscheduled, and 2) scheduled gates.

The difference between the two gate types lies in the temporal information about their execution. This information is contained only by the scheduled gates. In contrast, unscheduled gates are \emph{relatively} placed, and the following line is a string representing such a gate.
\begin{verbatim}
cx 4 0|0
\end{verbatim}

An unscheduled gate is not expected to appear at a certain time coordinate along the time axis of the quantum circuit (and for that reason in the corresponding \texttt{scqc}, too). The zero after the $|$ symbol means that the position of the $cx$ gate should not be offset in any time direction (past or future) after it is scheduled.

Scheduled gates appear at a well specified time coordinate in the quantum circuit (e.g. see next line):
\begin{verbatim}
1@cx 4 0
\end{verbatim}
The previous line is the string representation of a $cx$ gate scheduled to be executed at the (exact) time coordinate $1$. If the unscheduled gate had included the prefix $|3$, the time coordinate of the scheduled gate would have been $4@$ (1+3=4, because the computed coordinate $1$ would have been modified by the specified offset $3$).

SurfBraid \emph{does not include a single time unit}. Different transformation components can use different types of time xis with various units. For example, at quantum circuit level time could represent the position (index) of the gate along the longest path in the circuit, while the unit could be one (or more) plumbing piece(s) for \texttt{scqc} layouts.

\subsection{Automatic and Manual Wire Reordering}
\label{sec:reord}

Wire ordering can influence the amount of resources required for a \texttt{scqc}. For this reason, SurfBraid includes a straightforward transformation algorithm for ordering quantum wires according to their first usage. It also includes the wire recycling algorithm presented in \cite{paler2016wire}.

Furthermore, the user can control wire ordering from the GUI: it is possible to specify by clicks which wires to swap. For example, it is straightforward to obtain the wire permutation $w_2, w_0, w_1$ from an initial ordering such as $w_0, w_1, w_2$ (where $w_i$ is the name of the wires). Manual wire reordering allows the user to examine visually different \texttt{scqc} layout possibilities. Additional wire operations can be introduced in the SurfBraid workflow through the plugin system.

\begin{figure}[h!]
    \includegraphics[width=0.9\textwidth]{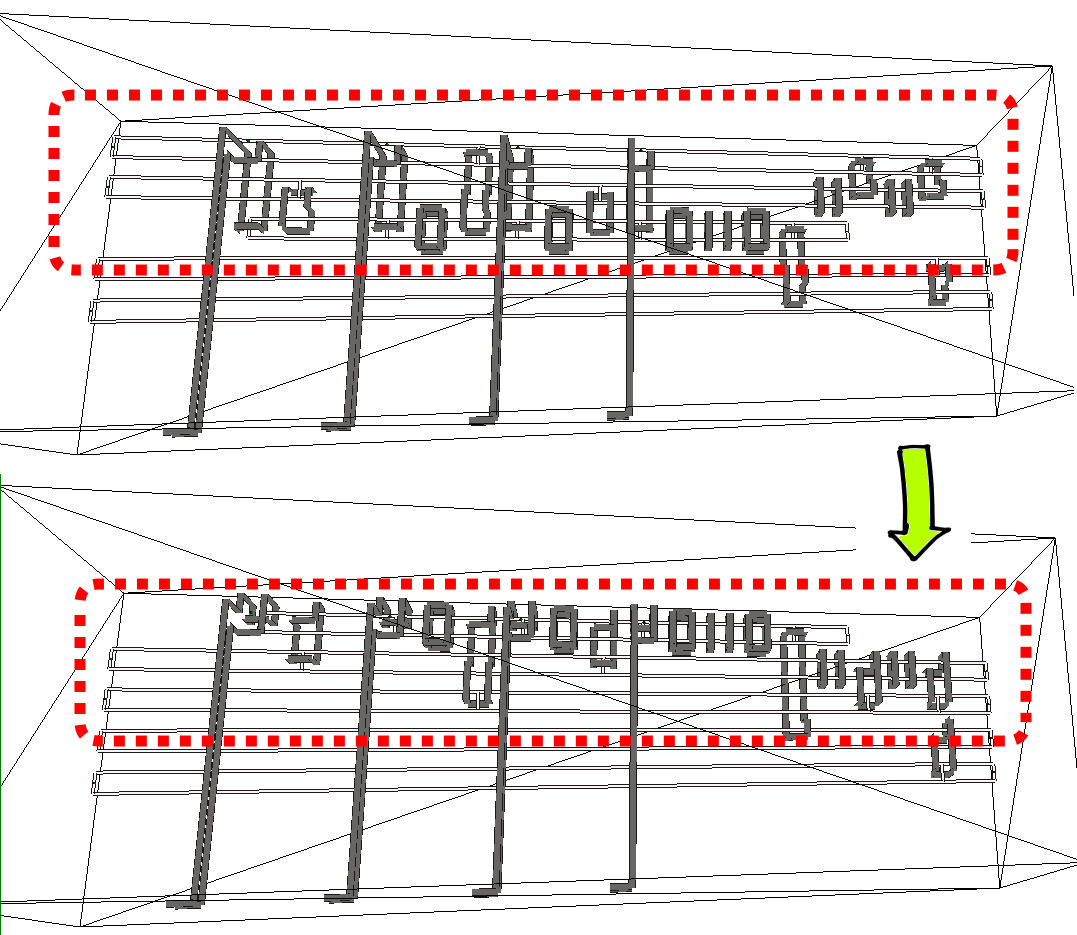}
    \caption{The result of reordering wires is visible when comparing the defects in the regions marked red. Time flows from the left to the right. Distillation boxes are not illustrated to simplify the view. The grey bounding box surrounding the defects indicates the total volume (in terms of plumbing pieces) when distillations are considered as part of the \texttt{scqc}.}
    \label{fig:reord}
\end{figure}

\subsection{Automatic Template Matching}
\label{sec:matching}

\begin{figure}[h!]
\centering
\includegraphics[width=0.9\columnwidth]{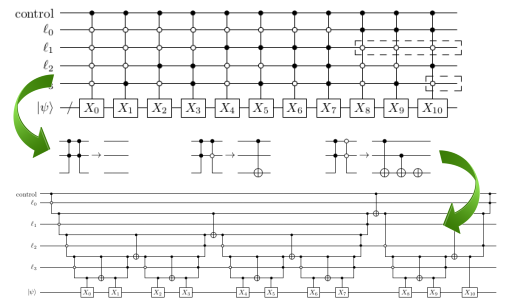}
\caption{Example of supported automatic transformation of the upper circuit into the lower one using the circuit rules illustrated in the middle \cite{babbush2018encoding}.}
\label{fig:tmpl}
\end{figure}

Gate lists are treated as simple string arrays, and this allows SurfBraid to easily support automatic decomposition of quantum gates. This implies that it was possible to implement the Clifford+T decompositions from \cite{paler2017fault} to generate fault-tolerant ICM representations of arbitrary quantum circuits. Automatic decompositions can be implemented one step further into automatic template matching, as exemplified in Figure~\ref{fig:tmpl}. SurfBraid includes this matching capability.

Template matching refers to applying local quantum circuit rewrite rules for optimisation purposes. A classic example is reducing the number of CNOTs in a circuit \cite{maslov2005quantum}. In general, the procedure is to consider each template application as a move in a higher dimensional search space, and the goal is related to a certain optimisation problem  (e.g. minimum number of CNOTs). Currently, SurfBraid supports only an exhaustive search strategy for a specified goal and a set of templates (rewrite rules). In Figure~\ref{fig:tmpl}, the rules are illustrated in the middle panel. The top panel circuit is sequentially transformed by each rule application until the circuit from the lower panel is generated. Thus, it can be stated that SurfBraid performs \emph{an exhaustive search by applying each template} until the optimal circuit (e.g. the gates have fewer controls) is found.

Matching is performed through transformation components, and the set of rewrite rules is arbitrary -- can be specified by the user (not from the GUI, for the moment). The search mechanism is intrinsic to the current prototype, but can be replaced through faster, but non-exact heuristic-based searches.

\subsection{T Gate Distribution Analysis}
\label{sec:distrib}

It was empirically shown in \cite{paler2018controlling} that, from the perspective of the resource estimation, the arrangement of the distillation procedures in the \texttt{scqc} layout is influenced by the temporal ordering of the T gates within the original Clifford+T circuit. In other words, the distribution of T gates in the circuit influences the \texttt{scqc} layout with respect to the placement of the distillation sub-circuits (green boxes in Figure~\ref{fig:qcirc}b).

Consequently, preparing and resource estimating \texttt{scqc} benefit from the analysis of Clifford+T circuits. Such an analysis has to be preceded by the execution of a transformation component for scheduling quantum gates. This is necessary for the gates to have known (exact) time coordinates. SurfBraid performs such circuit analysis through an algorithm parameterised by a look-ahead value: how many T gates does the circuit include in a given time window? The specified time window is calculated from the time coordinates of scheduled quantum gates (i.e. Section~\ref{sec:gatetypes}).

The \emph{optimal duration} of the time window is determined by the amount of hardware considered available for the \texttt{scqc}. For example, in a hardware restricted scenario, a single distillation box can be scheduled and executed at a time. The time window is dictated by the duration of the distillation procedure, and a single T gate would ideally be scheduled in that time window. If multiple T gates are detected in the same time window, all except one will be delayed (see next subsection). Each analysis will trigger the execution of a subsequent scheduling component which will refine temporal ordering of the already scheduled gates.

\begin{figure*}[h!]
\centering
\includegraphics[width=0.9\textwidth]{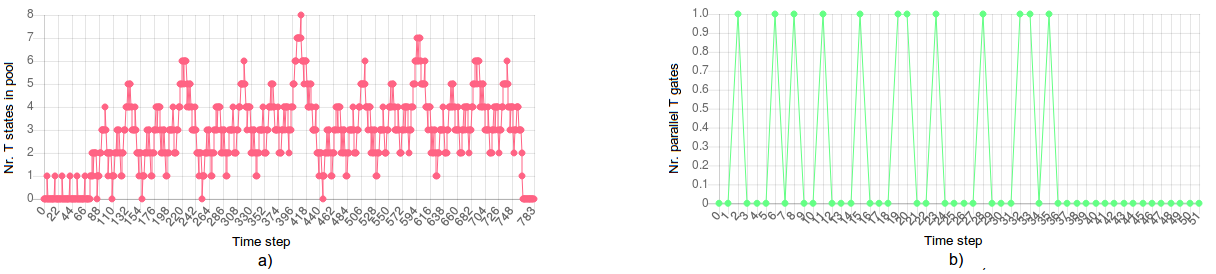}
\caption{Analysis diagrams computed automatically by SurfBraid: a) Number of T states available to the \texttt{scqc} computation; b) Distribution of T gates in the Clifford+T circuit \cite{paler2018controlling}.}
\label{fig:diag}
\end{figure*}


\subsection{Manual Control of Distillation Schedulers}
\label{sec:control}

There is a difference between scheduled gates (Section~\ref{sec:gatetypes}) and scheduled distillation boxes. The first refer to circuit-level operations, whereas the latter to structures abstracting the distillation sub-circuits from the \texttt{scqc} layouts. Distillation sub-circuits (boxes) are placed in a three-dimensional \texttt{scqc} layout by \emph{schedulers}. A scheduler determines the time coordinate when a distillation has to be started such that an error corrected computation can be executed without interruption.

SurfBraid allows the manual/automatic placement (scheduling) of boxes, because it is based on the software presented in \cite{paler2017online} and \cite{paler2017synthesis}. The complex scheduler from \cite{paler2017synthesis} has been used in a simplified form in \cite{paler2018controlling}, and only the latter can be directly used in SurfBraid. This scheduler variant assumes that \texttt{scqc} preparation is performed for a hardware-restricted environment: a single distillation can be placed at a time in the layout. The distillation schedulers are controlled by the analysis component for T state distillation distribution. This is a case of an analysis component steering the execution of a transformation component.

\section{Conclusion}

The concept (alpha version) of SurfBraid was introduced. The methodological novelty is the ability to automatically perform resource estimation for braided geometries of surface code protected quantum circuits. The tool can take arbitrary quantum circuits (or generated them if the corresponding generator script is programmed or imported), compile, and optimise the resulting \texttt{scqc}.

Technically future work will focus on: improving the GUI, eliminating software glitches, and unlocking more of the features of the C++ code into the Emscripten compiled module. The latter objective will increase the flexibility of the interactivity, because the vast majority of the scheduling control (as mentioned in Section~\ref{sec:control}) resides in the original C++ code.

From a surface QECC methodological perspective, the most important features are the following two.

\subsection{Supporting Lattice Surgery}

Lattice surgery \cite{litinski2018game} is gradually being considered a more resource efficient alternative to braided layouts of \texttt{scqc}. The present version of SurfBraid is operating only on braided layouts. The main application of SurfBraid is resource estimation, and it is reasonable to expect that there may be cases where surgery and braiding can complement each other. Therefore, the next versions will support lattice surgery for the beginning. Afterwards, the focus will be to enable a fair comparison between surgery and braiding.

\subsection{Mapping Circuits to Hardware}

Certain large scale quantum computing architectures will not operate on perfect lattices of physiscal qubits. Such lattices are necessary to enable the surface code error correction. Nevertheless, there are methods to renormalise faulty lattices \cite{herr2018local}, and it is of interest how this renormalisation affects the higher layer \texttt{scqc} layout.

\section*{Acknowledgement}
The author is grateful to Austin Fowler for his feedback and support, and to Craig Gidney for compiling a custom version of Quirk to be used in this tool. Simon J. Devitt's feedback on the alpha version was the impulse for finally submitting this work. This work was funded by project CHARON hosted by Linz Institute of Technology.

\bibliographystyle{IEEEtran}
\bibliography{surfbraid}

\begin{thebibliography}{10}
\providecommand{\url}[1]{#1}
\csname url@samestyle\endcsname
\providecommand{\newblock}{\relax}
\providecommand{\bibinfo}[2]{#2}
\providecommand{\BIBentrySTDinterwordspacing}{\spaceskip=0pt\relax}
\providecommand{\BIBentryALTinterwordstretchfactor}{4}
\providecommand{\BIBentryALTinterwordspacing}{\spaceskip=\fontdimen2\font plus
\BIBentryALTinterwordstretchfactor\fontdimen3\font minus
  \fontdimen4\font\relax}
\providecommand{\BIBforeignlanguage}[2]{{%
\expandafter\ifx\csname l@#1\endcsname\relax
\typeout{** WARNING: IEEEtran.bst: No hyphenation pattern has been}%
\typeout{** loaded for the language `#1'. Using the pattern for}%
\typeout{** the default language instead.}%
\else
\language=\csname l@#1\endcsname
\fi
#2}}
\providecommand{\BIBdecl}{\relax}
\BIBdecl

\bibitem{NC00}
M.~A. Nielsen and I.~L. Chuang, \emph{Quantum computation and quantum
  information}.\hskip 1em plus 0.5em minus 0.4em\relax Cambridge university
  press, 2010.

\bibitem{FMM13}
A.~Fowler, M.~Mariantoni, J.~Martinis, and A.~Cleland, ``{Surface Codes,
  Towards practical large-scale quantum computation},'' \emph{Phys. Rev. A.},
  vol.~86, p. 032324, 2012.

\bibitem{paler2017fault}
A.~Paler, I.~Polian, K.~Nemoto, and S.~J. Devitt, ``Fault-tolerant, high-level
  quantum circuits: form, compilation and description,'' \emph{Quantum Science
  and Technology}, vol.~2, no.~2, p. 025003, 2017.

\bibitem{raussendorf2007topological}
R.~Raussendorf, J.~Harrington, and K.~Goyal, ``Topological fault-tolerance in
  cluster state quantum computation,'' \emph{New Journal of Physics}, vol.~9,
  no.~6, p. 199, 2007.

\bibitem{litinski2018game}
D.~Litinski, ``A game of surface codes: Large-scale quantum computing with
  lattice surgery,'' \emph{arXiv preprint arXiv:1808.02892}, 2018.

\bibitem{van2016path}
R.~Van~Meter and S.~J. Devitt, ``The path to scalable distributed quantum
  computing,'' \emph{Computer}, vol.~49, no.~9, pp. 31--42, 2016.

\bibitem{steiger2018projectq}
D.~S. Steiger, T.~H{\"a}ner, and M.~Troyer, ``Projectq: an open source software
  framework for quantum computing,'' \emph{Quantum}, vol.~2, p.~49, 2018.

\bibitem{cross2018ibm}
A.~Cross, ``The ibm q experience and qiskit open-source quantum computing
  software,'' \emph{Bulletin of the American Physical Society}, 2018.

\bibitem{roetteler2017design}
M.~Roetteler, K.~M. Svore, D.~Wecker, and N.~Wiebe, ``Design automation for
  quantum architectures,'' in \emph{Proceedings of the Conference on Design,
  Automation \& Test in Europe}.\hskip 1em plus 0.5em minus 0.4em\relax
  European Design and Automation Association, 2017, pp. 1312--1317.

\bibitem{preskill2018quantum}
J.~Preskill, ``Quantum computing in the nisq era and beyond,'' \emph{arXiv
  preprint arXiv:1801.00862}, 2018.

\bibitem{paler2017synthesis}
A.~Paler, A.~G. Fowler, and R.~Wille, ``Synthesis of arbitrary quantum circuits
  to topological assembly: Systematic, online and compact,'' \emph{Scientific
  Reports}, vol.~7, no.~1, p. 10414, 2017.

\bibitem{suchara2013qure}
M.~Suchara, J.~Kubiatowicz, A.~Faruque, F.~T. Chong, C.-Y. Lai, and G.~Paz,
  ``Qure: The quantum resource estimator toolbox,'' in \emph{Computer Design
  (ICCD), 2013 IEEE 31st International Conference on}.\hskip 1em plus 0.5em
  minus 0.4em\relax IEEE, 2013, pp. 419--426.

\bibitem{lin2015paqcs}
C.-C. Lin, S.~Sur-Kolay, and N.~K. Jha, ``Paqcs: Physical design-aware
  fault-tolerant quantum circuit synthesis,'' \emph{IEEE Transactions on Very
  Large Scale Integration (VLSI) Systems}, vol.~23, no.~7, pp. 1221--1234,
  2015.

\bibitem{fowler2013surface}
A.~G. Fowler, S.~J. Devitt, and C.~Jones, ``Surface code implementation of
  block code state distillation,'' \emph{Scientific reports}, vol.~3, p. 1939,
  2013.

\bibitem{javadi2017optimized}
A.~Javadi-Abhari, P.~Gokhale, A.~Holmes, D.~Franklin, K.~R. Brown,
  M.~Martonosi, and F.~T. Chong, ``Optimized surface code communication in
  superconducting quantum computers,'' in \emph{Proceedings of the 50th Annual
  IEEE/ACM International Symposium on Microarchitecture}.\hskip 1em plus 0.5em
  minus 0.4em\relax ACM, 2017, pp. 692--705.

\bibitem{ding2018magic}
Y.~Ding, A.~Holmes, A.~Javadi-Abhari, D.~Franklin, M.~Martonosi, and F.~Chong,
  ``Magic-state functional units: Mapping and scheduling multi-level
  distillation circuits for fault-tolerant quantum architectures,'' in
  \emph{2018 51st Annual IEEE/ACM International Symposium on Microarchitecture
  (MICRO)}.\hskip 1em plus 0.5em minus 0.4em\relax IEEE, 2018, pp. 828--840.

\bibitem{smilkov2019tensorflow}
D.~Smilkov, N.~Thorat, Y.~Assogba, A.~Yuan, N.~Kreeger, P.~Yu, K.~Zhang,
  S.~Cai, E.~Nielsen, D.~Soergel \emph{et~al.}, ``Tensorflow. js: Machine
  learning for the web and beyond,'' \emph{arXiv preprint arXiv:1901.05350},
  2019.

\bibitem{green2013quipper}
A.~S. Green, P.~L. Lumsdaine, N.~J. Ross, P.~Selinger, and B.~Valiron,
  ``Quipper: a scalable quantum programming language,'' in \emph{ACM SIGPLAN
  Notices}, vol.~48, no.~6.\hskip 1em plus 0.5em minus 0.4em\relax ACM, 2013,
  pp. 333--342.

\bibitem{wootton2017decodoku}
J.~Wootton, ``Decodoku: Quantum error rorrection as a simple puzzle game,'' in
  \emph{APS Meeting Abstracts}, 2017.

\bibitem{soeken2012revkit}
M.~Soeken, S.~Frehse, R.~Wille, and R.~Drechsler, ``Revkit: A toolkit for
  reversible circuit design.'' \emph{Multiple-Valued Logic and Soft Computing},
  vol.~18, no.~1, pp. 55--65, 2012.

\bibitem{paler2017online}
\BIBentryALTinterwordspacing
A.~Paler, A.~G. Fowler, and R.~Wille, ``Online scheduled execution of quantum
  circuits protected by surface codes,'' \emph{Quantum Info. Comput.}, vol.~17,
  no. 15-16, pp. 1335--1348, Dec. 2017. [Online]. Available:
  \url{http://dl.acm.org/citation.cfm?id=3179584.3179589}
\BIBentrySTDinterwordspacing

\bibitem{zulehner2017advanced}
A.~Zulehner and R.~Wille, ``Advanced simulation of quantum computations:
  Compact representation rather than hardware power,'' \emph{arXiv preprint
  arXiv:1707.00865}, 2017.

\bibitem{maslov2005quantum}
D.~Maslov, C.~Young, D.~M. Miller, and G.~W. Dueck, ``Quantum circuit
  simplification using templates,'' in \emph{Proceedings of the conference on
  Design, Automation and Test in Europe-Volume 2}.\hskip 1em plus 0.5em minus
  0.4em\relax IEEE Computer Society, 2005, pp. 1208--1213.

\bibitem{babbush2018encoding}
R.~Babbush, C.~Gidney, D.~W. Berry, N.~Wiebe, J.~McClean, A.~Paler, A.~Fowler,
  and H.~Neven, ``Encoding electronic spectra in quantum circuits with linear t
  complexity,'' \emph{arXiv preprint arXiv:1805.03662}, 2018.

\bibitem{paler2018faster}
A.~Paler, A.~Fowler, and R.~Wille, ``Faster manipulation of large quantum
  circuits using wire label reference diagrams,'' \emph{arXiv preprint
  arXiv:1811.06011}, 2018.

\bibitem{fowler2012bridge}
A.~G. Fowler and S.~J. Devitt, ``A bridge to lower overhead quantum
  computation,'' \emph{arXiv preprint arXiv:1209.0510}, 2012.

\bibitem{gidney2018efficient}
C.~Gidney and A.~G. Fowler, ``Efficient magic state factories with a catalyzed|
  ccz> to 2| t> transformation,'' \emph{arXiv preprint arXiv:1812.01238}, 2018.

\bibitem{paler2016wire}
A.~Paler, R.~Wille, and S.~J. Devitt, ``Wire recycling for quantum circuit
  optimization,'' \emph{Physical Review A}, vol.~94, no.~4, p. 042337, 2016.

\bibitem{paler2018controlling}
A.~Paler, ``Controlling distilleries in fault-tolerant quantum circuits:
  problem statement and analysis towards a solution,'' in \emph{Nanoscale
  Architectures (NANOARCH), 2018 IEEE/ACM International Symposium on}.\hskip
  1em plus 0.5em minus 0.4em\relax IEEE, 2018.

\bibitem{herr2018local}
D.~Herr, A.~Paler, S.~J. Devitt, and F.~Nori, ``A local and scalable lattice
  renormalization method for ballistic quantum computation,'' \emph{npj Quantum
  Information}, vol.~4, no.~1, p.~27, 2018.

\end{thebibliography}

\end{document}